\documentclass[11pt]{amsart}
\usepackage{geometry}                % See geometry.pdf to learn the layout options. There are lots.
\usepackage[parfill]{parskip}    % Activate to begin paragraphs with an empty line rather than an indent
\usepackage{graphicx}
\usepackage{amssymb}
\usepackage{epstopdf}
\title{Does buckling instability of the pseudopodium limit how well an amoeba can climb?}
\author{Sandip Ghosal\\
Department of Mechanical Engineering\\
McCormick School of Engineering and Applied Science\\
Northwestern University\\
2145 Sheridan Road, Evanston, IL 60208\\
E-mail: s-ghosal@northwestern.edu\\
\\
Yoshio Fukui\\
Department of Cell and Molecular Biology\\
Feinberg School of Medicine\\
Northwestern University\\
303 East Chicago Avenue, Chicago, IL 60611\\
E-mail: y-fukui@northwestern.edu}
\begin{document}
\maketitle

\begin{abstract}
The maximum force that a crawling cell can exert on a substrate is a quantity  of interest 
in cell biomechanics. One way of quantifying this force is to allow the cell to crawl against
a measurable and adjustable restraining force until the cell  is 
no longer able to move in a direction opposite to the applied force. 
Fukui {\it et al.}~\cite{fukui_well_2000} reported on an experiment where 
amoeboid cells were imaged while they crawled against an artificial gravity field created by a centrifuge.
An unexpected observation was that the net applied force on the amoeba did not seem to be the primary 
factor that limited its ability to climb. Instead,  it appeared that the amoeba stalled 
when it was no longer able to support a pseudopodium against the applied gravity field. The high g-load 
bend the pseudopodium thereby preventing its attachment to the target point directly ahead of the cell. 
In this paper we further refine this idea by identifying the bending of the pseudopodium with the onset of 
elastic instability of a beam under its own weight. It is shown that the principal features of the experiment may 
be understood through this model and an estimate for the limiting g-load in reasonable accord with the experimental 
measurements is recovered.
\end{abstract}

%\pacs{87.17.Jj, 87.17.Rt, 87.17Aa}% PACS, the Physics and Astronomy
                             % Classification Scheme.
%\keywords{cell migration, centrifuge microscope, Dictyostelium discoideum, buckling instability, Euler-Bernoulli theory, turgor pressure}
%Use showkeys class option if keyword display desired
\maketitle
%\subsection{}
Motility is a fundamental trait that distinguishes living things. The macroscopic motion of plants and animals can 
ultimately be reduced to motion on the level of single cells~\cite{bray_cell_1992}. The movement of single cells also play a pivotal 
role in phenomena not directly related to movement of the whole organism. Thus, the morphogenetic migration of 
cells is responsible for the appearance of form and structure in embryogenesis. The movement of cells 
is crucial in the process of wound healing and in the functioning of the  immune system and unfortunately 
in metastasis, where cancer cells spread from the primary tumor to invade other organs of the body~\cite{li_biochemistry_2005}. 
When the environment of the cell is a fluid, the cell swims, but on a substrate or in a fibrous network,  
cells crawl. 
The mechanics of crawling~\cite{fletcher_introduction_2004,stossel_crawling_1993,stossel_machinery_1994,fukui_towardnew_1993,keller_falkovitz_83}
is a fundamental problem in biomechanics that has not yet been fully understood. 

The measurement of the actual forces exerted by crawling cells on substrates is of obvious importance and 
various techniques have been employed in order to perform such measurements~\cite{oliver_forces_1994}. One method is to restrain 
the cell in some way, for example by holding a micro-needle in its path or by applying a suction force 
with a micro-pippette. An ingenious non-invasive technique involves imaging the wrinkles on an elastic 
substrate from which the applied force may be inferred~\cite{dembo_stresses_1999}. 
Restraining forces can be applied to 
crawling cells by attaching magnetic beads to them and pulling with magnetic fields. Gravity is an excellent 
candidate for an external force as it does not require anything to be attached to the cell and it is the least likely to 
affect the natural behavior of the cell. 
Fukui {\it et al.}~\cite{fukui_well_2000} reported on an experiment in which amoebae of Dictyostelium
discoideum were allowed to crawl against an artificial gravity field created by a centrifuge. 
They determined  the maximum g-force 
at which the amoeba ``stalled'' -- that is, was unable to crawl in a direction opposite to the gravity field. However, a rather 
surprising finding was that the net force alone did not limit the crawling ability of the amoeba. 
Indeed, when the culture medium was replaced by one of a higher density, so that the amoeba was actually buoyant in it, it nevertheless 
stalled, even though, in this case, the gravity field was pushing the cell in the direction of motion. Fukui {\it et al.} observed 
that what appeared to limit the crawling ability of the amoeba was not the net force on the cell but rather the inability 
of the cell to extend a pseudopodium against the large gravitational field generated by the centrifuge; the pseudopodium was 
observed to bend and therefore not able to attach to a point on the substrate directly ahead of the amoeba
(see Figure~1). In the present study, we show that this proposal of Fukui {\it et al.} that the buckling instability of the pseudopodium is 
the major factor limiting the ability of cells to crawl against high gravitational forces, is supported by a quantitative 
analysis based on mechanics. We further suggest that the ability to crawl against high gravitational forces is
significantly improved in cells that are able to actively generate an internal turgor pressure in the pseudopodium by virtue of the contractile forces
in the posterior part of the cell cortex~\cite{fukui_towardnew_1993}.

\begin{figure}
\includegraphics[width = 0.75\textwidth,angle=0]{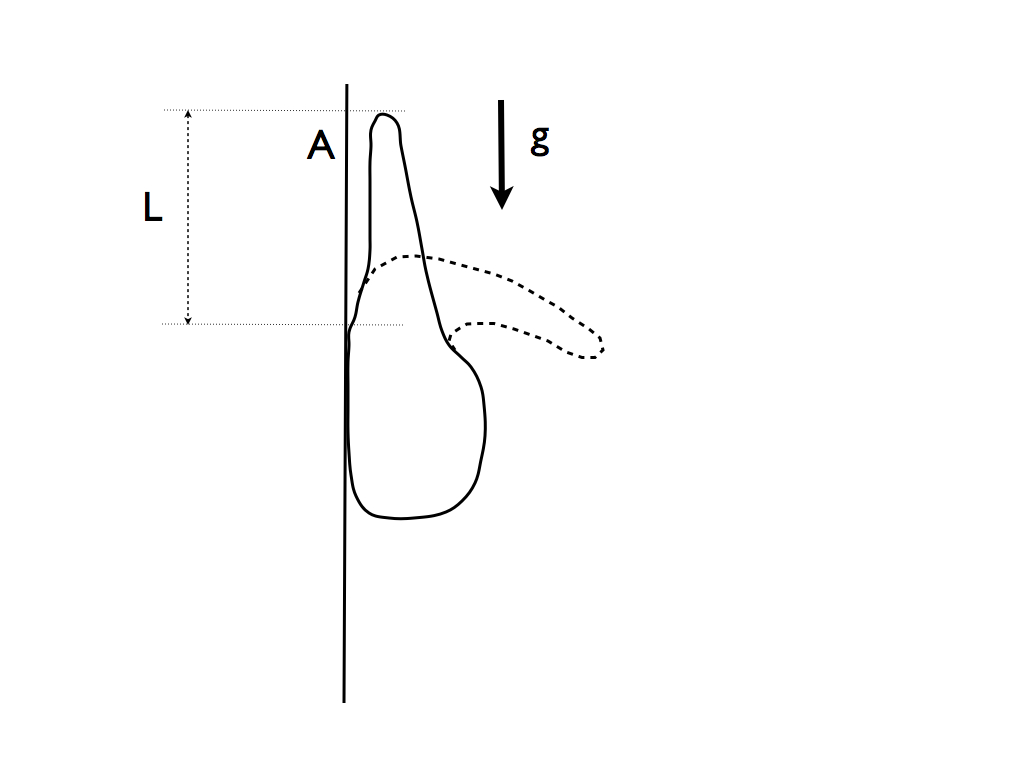}% Here is how to import EPS art
\caption{The amoeba stalls (is unable to move ``up'' against the artificial gravity field $g$ of the centrifuge) 
when the pseudopodium advancing in the direction of intended motion is bent by gravitational forces. Thus, the pseudopodium fails 
to attach at the target location A on the substrate directly above the cell body. Downward or lateral crawling of the cell 
is not prevented.}
\end{figure}
If one assumes, based on these observations, that the structural stability of the actin rich pseudopodium against the gravity 
field limits the crawling ability of the cell, then it should be possible to use classical results on the stability of elastic structures 
under gravity to estimate the stalling acceleration ($g_c$). 
The relevant result is the classical Euler-Bernoulli theory which  predicts
that an elastic beam of uniform cross-section 
buckles under a compressive force ($F$) if this force exceeds a critical value given by 
\begin{equation} 
F  = \frac{\alpha E I}{L^{2}},
\label{Euler-Bernoulli}
\end{equation}
where $E$ is the Young's modulus of the material, $I$ is the area moment of inertia of the cross-section,
$L$ is its length and the value of the numerical constant $\alpha$ depends on the conditions at the ends of the beam;
for a beam clamped at one end and free at the other, $\alpha = \pi^{2}/4$.
The Euler-Bernoulli theory proceeds from the assumptions of mechanical equilibrium and small deformations that result in a linear boundary value problem for the beam centerline.
The requirement that this equation admit nonzero solutions results in an eigenvalue problem, and 
Eq.~\ref{Euler-Bernoulli}, corresponds to the  lowest eigenvalue~\cite{timoshenko_theory_2009}. 
If the compressive force is the weight of the 
beam ($W$), a simple estimate for the maximum weight up to  which the vertical configuration could be stable 
may be obtained by assuming all of the weight to be concentrated at the center of mass, so that 
\begin{equation} 
W  = \frac{4 \alpha E I}{L^{2}}.
\label{crit_W}
\end{equation}
A more careful analysis where the weight is assumed to be uniformly distributed along the beam was provided by Greenhill~\cite{a.g._greenhill_determination_1881}.
Eq.~\ref{crit_W} was found to hold except that  $\alpha \approx 1.99$. For tapered beams 
Eq.~\ref{crit_W} may still be applied if $I$ is regarded as the area moment of inertia of the base. Greenhill showed 
that for a right circular cone $\alpha \approx 2.54$ and for a paraboloid of revolution $\alpha \approx 2.47$. 
Keller and Niordson~\cite{keller_tallest_1966} calculated the greatest height $L$ that a beam of fixed weight $W$ can have if it is allowed to taper in an arbitrary way while preserving the shape of the cross-section. 
Once again, the result can be expressed in the form of Eq.~\ref{crit_W} with $\alpha \approx 8.23$. Thus, 
Eq.~\ref{crit_W}, where $\alpha$ is a numerical constant that has a value roughly between one and ten, 
may be used to estimate the critical acceleration $g_c$ of the centrifuge.
For this purpose, one can rewrite Eq.~\ref{crit_W} as 
\begin{equation} 
g_c  = \frac{4 \alpha E I}{(\Delta \rho) V L^{2}},
\label{crit_g}
\end{equation}
using  the apparent weight~\cite{kokkinis_effect_1987} (in place of $W$) of the pseudopodium which is the difference in density between the pseudopodium and the culture medium ($\Delta \rho$) 
multiplied by the volume ($V$) of the pseudopodium\footnote{this correction in the formula for the stability threshold is necessary  because the equation 
for the deformation $y(x)$ of the beam is obtained by minimizing the energy functional $\int_{0}^{L} dx \left[ EI (y^{\prime \prime})^{2}/2 + (\rho - \rho_{0}) g \delta X   \right]$
where $\delta X$ is the vertical  displacement of a point on the beam at location `$x$' due to beam curvature and $\rho_0$ is the density of the external 
medium. This differs from the corresponding expression in the absence of the external medium in that the density difference $\Delta \rho = \rho - \rho_0$ 
replaces the density of the material of the beam, $\rho$.}.

If the pseudopodium is regarded as a circular cylinder 
of diameter $2$ $\mu$m and length $5$ $\mu$m, then $I \approx 0.8 \times 10^{-24}$ m$^{4}$ and 
$V \approx 1.6 \times 10^{-17}$ m$^{3}$. Fukui {\it et} al~\cite{fukui_well_2000} estimate that the density of the actin 
rich pseudopodium must be at least $\rho = 1.124$ gm/cm$^{3}$. If we accept this value, then, since the medium density (at 0 \% Percoll) 
is $\rho_0 = 1.005$ gm/cm$^{3}$, $\Delta \rho = \rho - \rho_0 = 0.119$ gm/cm$^{3}$. 
The greatest uncertainty arises in estimating the Young's modulus 
$E$. If the pseudopodium is presumed to be supported predominantly by the mechanical 
strength of the actin network one could use in vitro measurements of the strength of actin gels. Such measurements 
are usually expressed in terms of the shear modulus $G$ which is related to $E$ and the bulk modulus $K_V$ as 
$E = 9 G K_{V}/(3 K_{V} + G) \approx 3 G$ (since $G/K_{V}$ is smaller than $10^{-7}$). In vitro 
measurements show that $G$ for actin networks  vary from about $300$ Pa in the ``gel'' state to a
 value three orders of magnitude lower~\cite{janmey_mechanical_1994,gardel_elastic_2004} in the fluid state.
The measured value depends primarily on the 
density and length of actin filaments and the density of cross-links created by various actin binding proteins (ABPs). If we take 
$G \approx 300$ Pa and $\alpha \approx 2.5$ (corresponding to a structure of paraboloid shape) Eq.~\ref{crit_g} yields a numerical estimate $g_c \sim 5  \times 10^{4}$ m/s$^{2}$ .

In the experiment~\cite{fukui_well_2000}  Fukui {\it et al.} found that the myosin II knockout  mutant (HS1) of the Dictyostelium amoeba 
stalled at $g_c \approx 3.9 \times 10^{4}$ m/s$^{2}$ (in the buffer without Percoll) which is in reasonable accord with the above
 estimate. The wild type strain (NC4) containing myosin II did not stall even at the highest accelerations tested 
(about $11.2 \times 10^{4}$  m/s$^{2}$). This is probably because the wild type cells are able to create considerable 
turgor pressure due to the myosin II-dependent contractile forces in the actin cortex thereby stiffening the 
pseudopodium. This case is discussed next.

\begin{figure}
\includegraphics[width = 0.75\textwidth,angle=0]{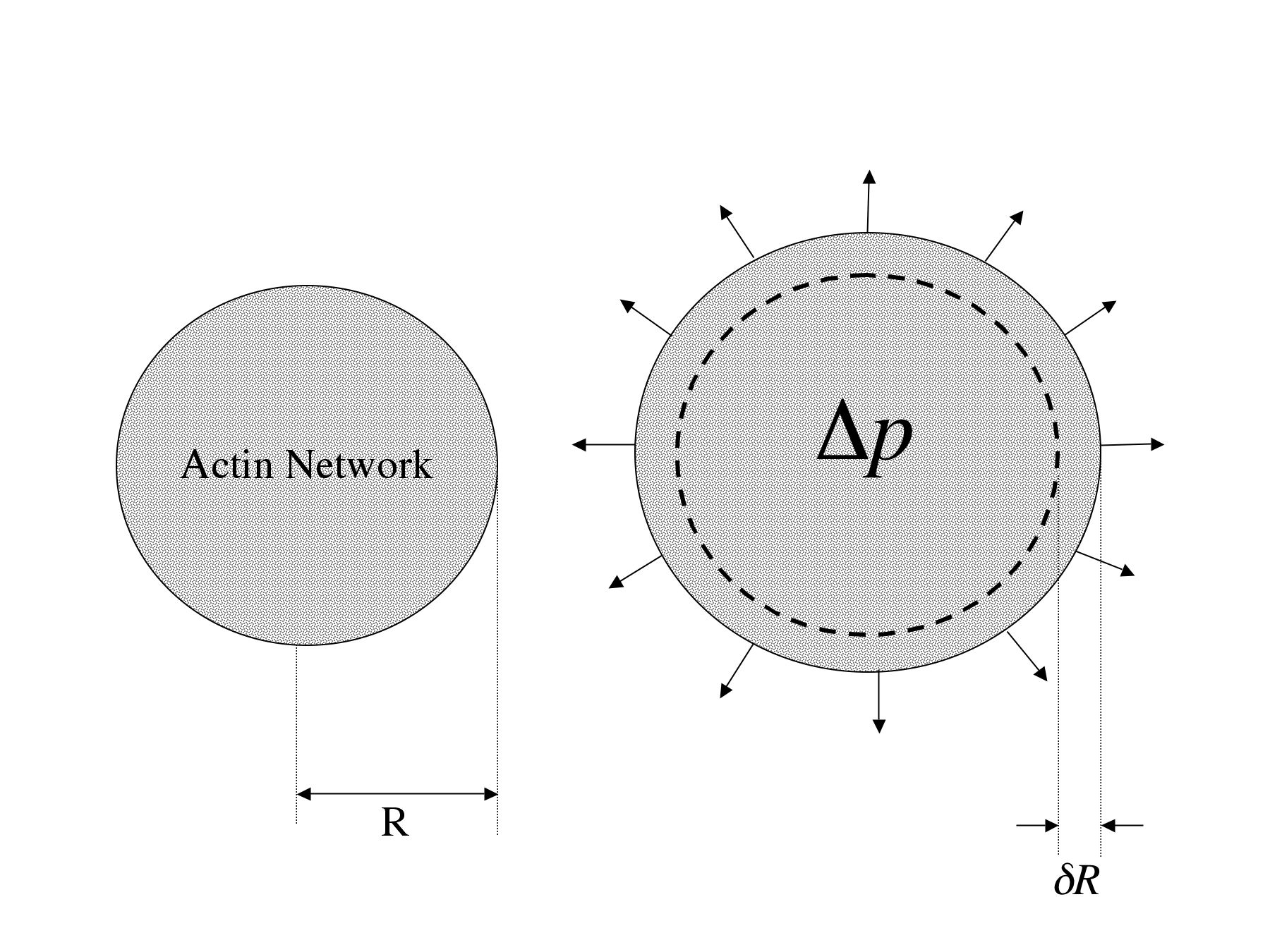}% Here is how to import EPS art
\caption{Diagram illustrating the radial expansion of the pseudopodium as a consequence 
of the rise in internal turgor pressure, resulting in a ``stiffening'' of the structure against 
elastic buckling.}
\end{figure}

In order to understand the effect of turgor pressure one needs to recognize that the pseudopodium is a poroelastic solid.
When a beam made of such a material bends, the compression of the pores on one side of the neutral line results in a pressure 
rise in the interstitial fluid which then drives a flow across the beam. On account of viscous resistance, the fluid 
responds to the bend with a time lag so that the system behaves much like a coupled mass-spring-damper system~\cite{maha_2004}.
However,  this mechanism does not alter the stability limit since  the bifurcation at the onset of elastic instability takes place at zero frequency.
Internal hydrostatic pressure stiffens the structure by one of two mechanisms (a) the swelling and consequent 
stretching of elastic elements may put it in a regime where the stress  strain relation is no longer linear 
(b) the swelling may alter geometrical parameters (specifically, the parameter $I$ in Eq.~\ref{crit_g}).
The first of these effects is less likely, though  a careful estimate is difficult as not much information is available on the nonlinear elasticity 
of actin gels. However, the second of these effects is readily estimated. The situation is depicted in Figure~2. 
Since an element of the cell membrane is in equilibrium due to the balance of an outward pressure ($\Delta p$) and 
an inward elastic stress $E (\delta R / R)$ where $\delta R$ is the increase in the radius $R$, we have 
$\delta R/R = (\Delta p) / E$. Thus, 
\begin{equation} 
1 + \frac{\delta g_c}{g_c} = \left( 1 + \frac{\delta R}{R}  \right)^{4} =  \left( 1 + \frac{\Delta p}{E}  \right)^{4}
\end{equation}
since the area moment of inertia of a cylinder is $I = \pi R^{4} /4$.
In the experiment~\cite{fukui_well_2000}  the  wild type strain (NC4) containing myosin II did not stall at
the maximum acceleration of $g_c = 11.2 \times 10^{4}$  m/s$^{2}$, suggesting that $\delta g_c  / g_c > 1.9$.
Thus, $\Delta p > 0.305 \times E = 275$ Pa, using the value $E=3G=900$ Pa cited earlier.  Pasternak {\it et al.}~\cite{pasternak_capping_1989}
 report a difference $\Delta T \approx 0.13$ mdyn per 
 micron in the cortical tension between the strains AX4 and the myosin lacking mhcA- strain of Dictyostelium
 both in the resting phase. If one converts this number to a pressure using the 
 Laplace formula for surface tension, one obtains the estimate for the myosin generated pressure: 
 $\Delta p = 2 \Delta T / R \sim 520$ Pa where $R \sim 5$ $\mu$m 
 is taken as a characteristic radius of the cell. Thus, our rough estimate
 $\Delta p > 275$ Pa is not inconsistent with reported values for myosin II-dependent
 pressures in the cell cortex that may be inferred on the basis of existing experimental data. 
 
In conclusion, the hypothesis advanced by Fukui {\it et al.} that the ability of the Dictyostelium amoeba 
to support a pseudopodium against a strong  gravity field limits its ability to crawl against such a field 
appear to be supported by the present analysis based on the mechanics of the buckling of elastic structures. Similar 
ideas have long been used in the field of plant biomechanics~\cite{vogel_living_2006}, but the centrifuge 
experiments of Fukui {\it et al.} present an opportunity for an application of these concepts to 
the mechanics of cell crawling. In a broader context, we illustrate that the interpretation of experiments 
in which one attempts to quantify the force applied by a crawling cell on a substrate by measuring 
an applied restraining force~\cite{oliver_forces_1994} may be subtle, because, the limiting factor may not be the ability of the 
cell to pull against the applied force but rather a failure in some other aspect of the motility mechanism
of the cell (e.g. contact inhibition~\cite{weiss_guiding_1961}).\\[2ex]

\noindent {\it Acknowledgements:} We would like to thank Howard A. Stone and Joseph B. Keller for reading and commenting on a draft of the 
manuscript and L. Mahadevan for helpful discussions relating to the bending of poroelastic beams.
\bibliographystyle{prsty}
%\bibliography{zotero_lib}

\end{document}